\documentclass[11pt,epsfig]{article}
\usepackage{moriond,epsfig}

\def\be{\begin{equation}}
\def\ee{\end{equation}}
\def\bea{\begin{eqnarray}}
\def\eea{\end{eqnarray}}

\begin{document}
\begin{flushright}
IPM/P-2005/033 \\
hep-ph/0505004 \\
\today
\end{flushright}
\vspace*{4cm}
\title{ON EFFECTS OF THE LARGE NEUTRINO B-TERM ON  LOW
ENERGY PHYSICS}

\author{ YASAMAN FARZAN }
\address{Institute for Studies in Theoretical Physics and Mathematics
(IPM)\\
P. O. Box 19395-5531, Tehran, Iran}

\maketitle\abstracts{ To embed the seesaw mechanism in the MSSM,
two or three right-handed neutrino supermultiplets, $N_i$, have to
be added to the model. In this framework, the supersymmetry
breaking potential will include a new term called neutrino
$B$-term: $MB_\nu \tilde{N} \tilde{N}/2$. In this talk, we present
a toy model that generates a large neutrino $B$-term keeping other
supersymmetry breaking parameters  small. We then review the
consequences of having a large neutrino $B$-term on the
electroweak symmetry breaking parameters and electric dipole
moments (EDMs) of elementary particles.}

\section{Introduction}
The Standard Model (SM) of elementary particles has been so far
able to explain the accelerator data. Despite its remarkable
achievements, the SM suffers from some shortcomings: i) In order
to cancel the quadratic divergences appearing in the radiative
corrections to the Higgs mass, the parameters of the theory have
to be highly fine-tuned. (ii) In the framework of SM with zero
neutrino mass, we are unable to explain the solar and atmospheric
neutrino data. We also need physics beyond the SM to explain the
KamLAND and K2K results.

To overcome  the former shortcoming, several extensions  of the SM
have been developed among which the Minimal Supersymmetric
Standard Model (MSSM) is one of the most elegant  models. The
superpotential of this model is \be W=Y_\ell^{i j}
\epsilon_{\alpha \beta}H_d^\alpha E_{i}  L_j^\beta+\mu H_u H_d,
\ee where $L_j^\beta$ is the supermultiplet corresponding to the
 doublet $(\nu_{Lj},~ l_{Lj})$  and $E_i$ is the superfield associated with the charged lepton
 $\ell_{iL}^+$. $Y_\ell$ is the Yukawa matrix
 of the charged leptons and the last term is the famous mu-term.

The MSSM in its most general form contains several sources of
Lepton Flavor Violation (LFV) which in principle can give rise to
LFV rare decays of $\mu$ and $\tau$ exceeding the present
experimental bounds. Motivated by this observation, the
constrained MSSM (CMSSM) has been proposed which assumes that at
high energies, which we will loosely call $M_{GUT}$, the masses of
sfermions are universal. That is, at $M_{GUT}$, the soft
supersymmetry breaking potential is
 \bea
-{\cal L}_{soft} &=&
m_0^2(\tilde{L}_{L\alpha}^\dagger\tilde{L}_{L\alpha}+
\tilde{E}_{\alpha}^\dagger \tilde{E}_{\alpha}+H_d^\dagger H_d+
H_u^\dagger H_u \label{soft}) +
 \frac{1}{2} m_{1/2}(\tilde{B}^\dagger \tilde{B}+
\tilde{W^a}^\dagger \tilde{W^a}) \cr &+& (b^0_H  H_d H_u +h.c.)+
A_\ell^{ij}\epsilon_{\alpha \beta} H_d^\alpha\tilde{E}_{i}
\tilde{L}_{Lj}^\beta+{\rm terms \ involving \ quarks}, \eea where
the $A$-coupling is supposed to be proportional to  the
corresponding Yukawa couplings, $A_\ell=a_0 Y_\ell$.

Although there are strong upper bounds on neutrino mass from beta
decay experiments and cosmological considerations, the observation
of neutrino oscillation guarantees that at least two neutrinos
have nonzero mass. That is while in the framework of both SM and
MSSM the masses of neutrinos are zero. One of the most economic
ways to attribute a tiny but nonzero mass to neutrinos is the
famous seesaw mechanism which involves three very heavy
right-handed neutrinos. To embed the seesaw mechanism in the MSSM,
three right-handed neutrino supermultiplets, $N_i$, have to be
added to the model. In the presence of these new supermultiplets
the superpotential includes new terms $$\Delta W^N=Y_\nu^{i j}
\epsilon_{\alpha \beta}H_u^\alpha N_{i}
L_j^\beta+\frac{1}{2}M_{ij} N_{i} N_{j},$$ where the first term is
the Yukawa coupling of neutrinos and the second term is the mass
term for $N_i$. Without loss of generality we can rotate and
re-phase the fields to make both $Y_\ell$ and $M_{ij}$ real
diagonal. Throughout this paper, we work in such basis:
$Y_\ell^{ij}={\rm diag}(Y_e, Y_\mu, Y_\tau)$ and $M^{ij}={\rm
diag}(M_1, M_2, M_3)$. In order to make  neutrino masses tiny,
$M_i$ have to be very large: $M_i/M_{susy}\gg 1$. In the presence
of $\tilde{N}_i$, also the soft supersymmetry breaking potential
includes new terms: \be-\Delta {\cal L}_{soft}^N=m_0^2
\tilde{N}_i^\dagger \tilde{N}_i+ A_\nu^{ij}\epsilon_{\alpha\beta}
H_u^\alpha \tilde{N}_{i} \tilde{L}_{Lj}^\beta + (\frac{1}{2} B_\nu
M_i \tilde{N}^i\tilde{N}^i+h. c.), \ee where at the GUT scale
$A_\nu=a_0 Y_\nu$. The last term is the neutrino B-term which
violates the lepton number by two units. Since $\tilde{N}^i$ are
singlets of SU(3)$\times$ SU(2)$\times$ U(1), in general the
$B_\nu$ can be much higher than the electroweak scale, $m_{EW}$.
However for the range of parameters that $m_{EW}\ll B_\nu \ll
M_i$, the contribution of the neutrino B-term to neutrino masses
\cite{grossman}$^)$, electroweak symmetry breaking parameters
\cite{electroweak}$^)$ and LFV masses of the left-handed sleptons
\cite{bterm}$^)$ can be significant. Also, if $ B_\nu$ is complex,
it can be considered as a new source of CP-violation, inducing
EDMs for elementary particles \cite{bterm}$^)$.

This paper is organized as follows. In Sec. \ref{toy}, we review
the theoretical prediction for the order of magnitude of $B_\nu$
in the context of mSUGRA and we then  suggest a toy model that
allows large values of $B_\nu$  while keeping other supersymmetry
breaking parameters low ($\stackrel{<}{\sim} 1$ TeV). In Sec.
\ref{mHu}, we review the effects of $B_\nu$ on electroweak
symmetry breaking parameters. In Sec. \ref{edm}, we study the
effects of an imaginary $B_\nu$ on the EDMs of the elementary
particles. In Sec. \ref{conclusion}, we summarize our conclusions.
\section{Theoretical Expectation for $|B_\nu|$}\label{toy} In the
context of the mSUGRA, the soft supersymmetry breaking terms
originate from the interaction of a chiral superfield $S$ with the
super-potential: \be \int d^2 \theta S(\theta) W(\theta). \ee The
scalar and $F$-components of $S$ develop vacuum expectation values
$\langle S \rangle=1 +F_S \theta^2$ and $\langle F_S \rangle$
determines the scale of  the soft supersymmetry breaking terms.
Within this model we expect $B_\nu \sim a_0 \sim m_{susy}$.
Remember that we have parameterized the neutrino $B$-term as $M
B_\nu \tilde{N}\tilde{N}/2$ so, in this model, we expect
$\sqrt{B_\nu M}\gg m_{susy}$.

Let us now suppose that besides $S$ which couples to the lepton
number conserving part of the superpotential, there is  a spurion
field, $X$, that carries lepton number equal to two. We can then
write the following term in the superpotential \be
\label{spurion}\int d^2 \theta \lambda_i X N_i N_i. \ee However,
terms such as $\int d^2 \theta X H_u H_d$ are forbidden by lepton
number conservation. Moreover, terms such as $\int X^\dagger X
\Phi^\dagger \Phi d^4 \theta$ in the K\" ahler potential are
suppressed by powers of $M_{pl}^{-1}$. Let us assume that the
self-interaction of the hidden sector is such that both the
scalar- and $F$-components of $X$ develop nonzero vacuum
expectation values. The vacuum expectation values  of the
components of $X$ break the lepton number symmetry of the model.
The vacuum expectation value of the scalar component of $X$,
$\langle \tilde{X}\rangle$, corresponds to the Majorana mass term
of the right-handed neutrinos while the vacuum expectation value
of the $F$-component, $\langle F_X \rangle$, gives  the neutrino
$B$-term. With our parametrization of the neutrino $B$-term, \be
B_\nu ={\langle F_X\rangle \over \langle \tilde{X} \rangle}. \ee
Both $\langle F_X\rangle$ and $\langle \tilde{X} \rangle$ can be
large, giving rise to large right-handed neutrino masses and
$B_\nu$, while other supersymmetry breaking terms, which are given
by $\langle F_S \rangle$, are at the TeV scale or smaller. Notice
that in this model, in the basis that the mass matrix of the
right-handed neutrinos is real diagonal, the neutrino $B$-term is
also diagonal so the parametrization that we are using for the
neutrino $B$-term is the appropriate one.

\section{Effects of the Neutrino $B$-term on the Higgs Mass
Parameters}\label{mHu} In this section, we study the effects of a
large $B_\nu$ on the Higgs mass parameters and derive bounds on
its value from the fulfillment of the electroweak symmetry
breaking condition.

\begin{figure}
\psfig{figure=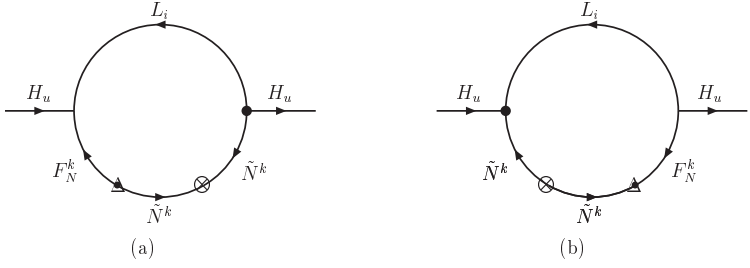,bb=75 578 528 744, clip=true, height=1.5in}
\caption{Diagrams contributing to $m_{H_u}^2$. $F_N^k$ represents
the auxiliary field associated with the right-handed neutrino,
$N_k$. The $A_\nu$ vertices are marked with black circles. The
neutrino $B$-term and $M$ insertions are indicated by $\otimes$
and $\Delta$, respectively.}
 \label{bh2}
\end{figure}

Diagrams shown in Fig. \ref{bh2} give a correction to $m_{H_u}^2$
which is equal to \be -i \Delta m_{H_u}^2=2\sum_k \int { M_k^2
{\rm Re}[B_\nu \sum_i (Y_\nu)_{ki} (A_\nu^*)_{ki}] \over
k^2(k^2-M_k^2)^2} {d^4 k \over (2\pi)^4} =-i 2\sum_{k,i} {\rm
Re}{\left[ B_\nu {\rm Tr}(Y_\nu A_\nu^\dagger)\right]\over 16
\pi^2}. \label{mhucorrection} \ee

\begin{figure}
\psfig{figure=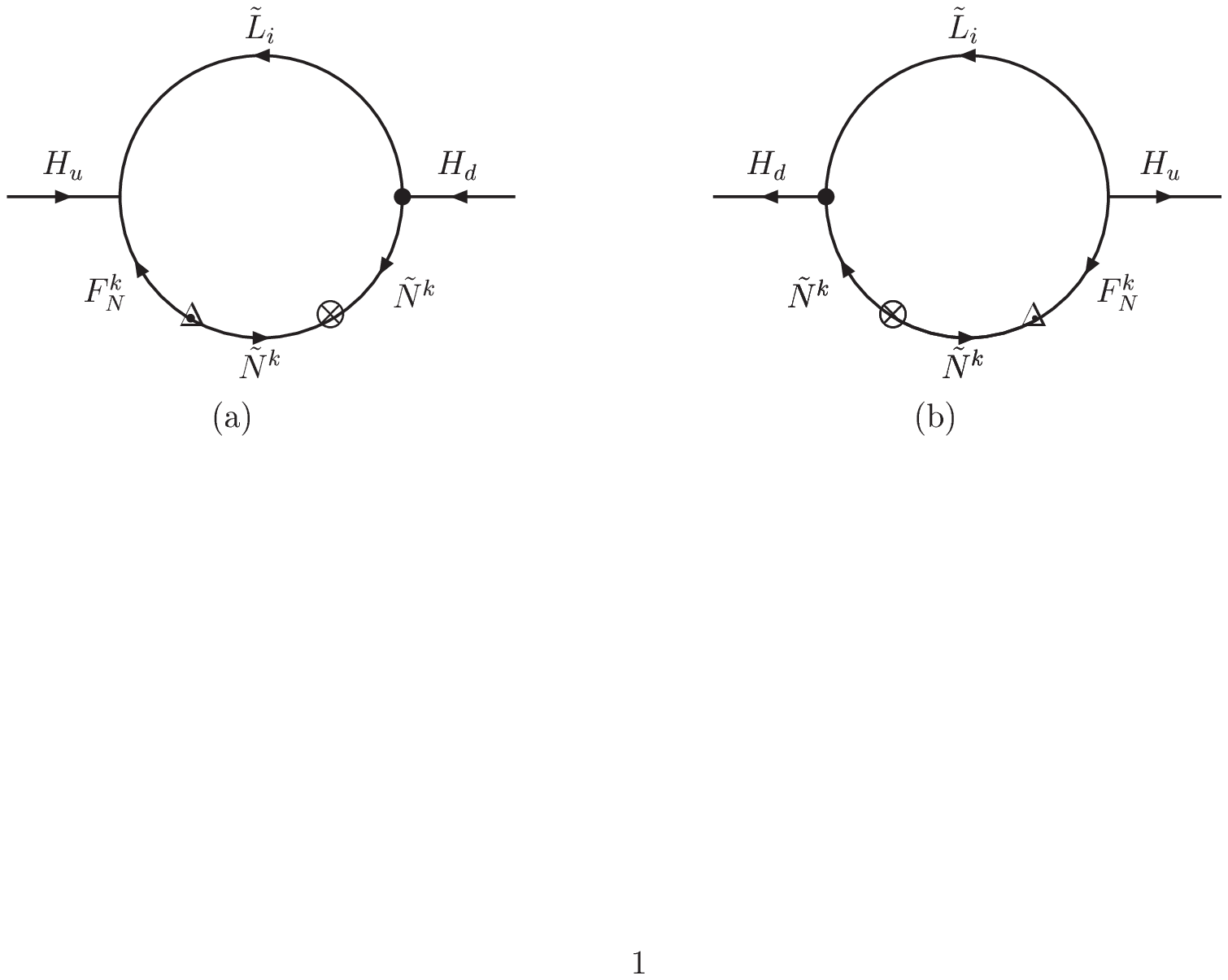,bb=68 273 522 445, clip=true, height=1.5in}
\caption{
 Diagrams contributing to
 the Higgs $B$-term. $F_N^k$ represents the auxiliary
field associated with the right-handed neutrino, $N_k$. The  two
vertices marked with black circles are given by $\mu
Y_\nu^\dagger$ and $\mu^* Y_\nu$. The neutrino $B$-term and $M$
insertions are indicated by $\otimes$ and $\Delta$, respectively.}
\label{bbh}
\end{figure}

Presence of a large neutrino $B$-term also induces non-negligible
corrections to $b_H$ as it is shown in Fig \ref{bbh}. The
correction is finite and is equal to \be \label{bhcorrection} -i
\Delta b_H=-B_\nu \sum_k \int {M_k^2 {\rm Tr}\left[ (Y_\nu)_{ki}
(Y_\nu^*)_{ki} \right] \mu \over k^2(k^2-M_k^2)^2} {d^4k \over (2
\pi)^4} ={i B_\nu \mu {\rm Tr}\left[ Y_\nu Y_\nu^\dagger\right]
\over (4\pi)^2}.\ee
By dimensional analysis we can show that any correction due to
$B_\nu$ to the quadratic Higgs interaction is suppressed by $B_\nu
/M$ which is negligible. The contribution to the cubic Higgs term
is also zero. So the  potential of $H_u^0$ and $H_d^0$ is  \be
V=(|\mu|^2+m_{H_u}^2 +\Delta
m_{H_u}^2)|H_u^0|^2+(|\mu|^2+m_{H_d}^2)|H_d^0|^2
 \label{V}
\ee
$$+[(b_H+\Delta b_H) H_u^0H_d^0+{\rm H. c.}]+{g^2+g'^2\over
8}(|H_u^0|^2-|H_d^0|^2 )^2.
$$
  Note that here we have not included the one-loop effective
  potential terms \cite{coleman}$^)$; however, since our analysis  is
  based on an order of magnitude consideration, including those terms
  cannot alter our conclusions.

Requiring $m_Z^2=(g^2+g'^2)(\langle H_u\rangle^2+\langle
H_d\rangle^2)/2$ and $\partial V/\partial H_u^0=\partial
V/\partial H_d^0=0,$ we find \be |\mu|^2+m_{H_d}^2=\left|
b_H+\Delta b_H\right| \tan \beta -(m_Z^2/2)\cos 2\beta\label{con1}
\ee and \be |\mu|^2+m_{H_u}^2+\Delta m_{H_u}^2=\left| b_H+\Delta
b_H \right| \cot \beta +(m_Z^2/2)\cos 2\beta\label{con2} \ee where
$\tan \beta =\langle H_u\rangle /\langle H_d \rangle $. Assuming
$|\mu|^2\sim m_{H_u}^2 \sim m_{susy}^2$, Eq. (\ref{con1}) gives
\be \label{bound} \left| b_H-B_\nu \mu {{\rm Tr} [Y_\nu
Y_\nu^\dagger ] \over 16 \pi^2} \right|\sim m_{susy}^2/\tan \beta.
\ee From the LEP data \cite{lep}$^)$, we know that $\tan \beta>2 $
and the data favors large values of $\tan \beta$ ($\tan \beta
>10$). Based on the naturalness condition, it seems quite unlikely
that $b_H$ and $\Delta b_H$ cancel each other, so we expect that
\be B_\nu Y_\nu^2/(16 \pi^2) <m_{susy}/\tan \beta. \label{dis} \ee
Notice that if $Y_\nu\ll 1$, $B_\nu$ can  still be several orders
of magnitude larger than $m_{susy}$.

\section{Effects of an Imaginary $B_\nu$ on Electric Dipole
Moments}\label{edm} In the CMSSM, in addition to the phase of the
CKM matrix, there are two sources of CP-violation which are
usually  attributed to the phases of $\mu$ and $a_0$. These two
phases can induce EDMs for the electron, neutron and mercury.
Combining the bounds on $d_e$ and $d_{Hg}$, one can derive strong
bounds on Im$[a_0]$ and $\phi_\mu$ \cite{nocancelation}$^)$.
Adding the three heavy right-handed neutrinos to the model new
sources of CP-violation emerge; six physical phases associated
with $Y_\nu$ and the phase of $B_\nu$. The effects of the phases
of $Y_\nu$ on the EDMs of charged leptons have been studied in a
series of papers \cite{ellis,michael}$^)$. The effects of an
imaginary $B_\nu$ on $d_e$ has been first noticed in Ref.
\cite{bterm}$^)$. In this section, we briefly review the latter
effect.

\begin{figure}
\psfig{figure=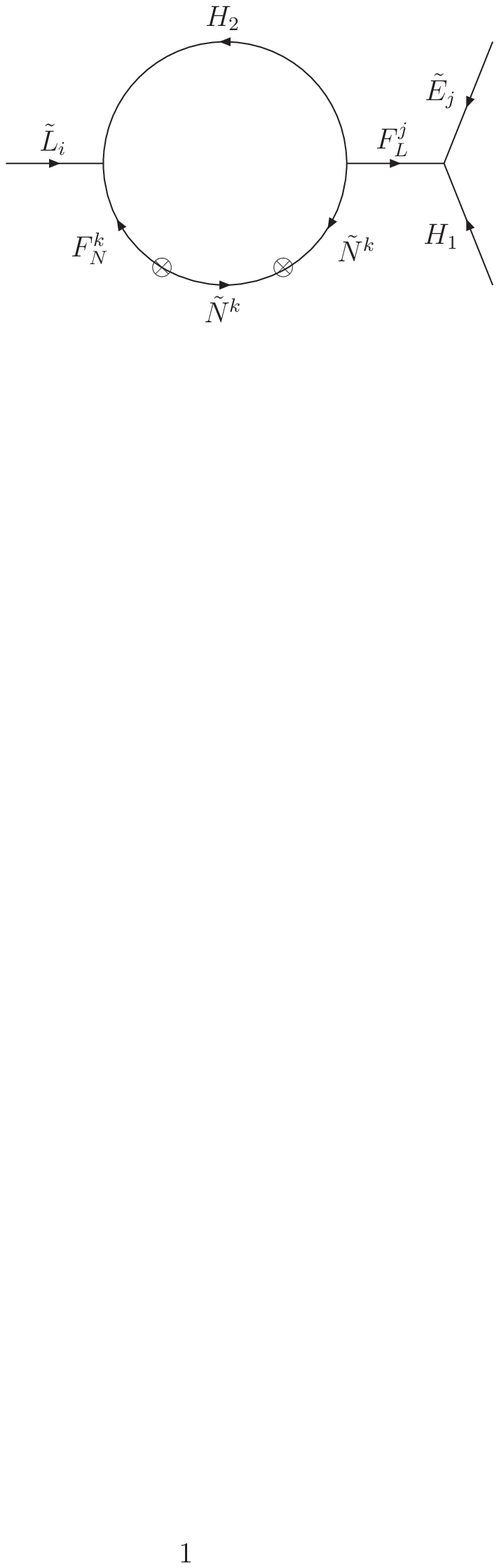,bb=219 533 450 695, clip=true, height=1.5in}
\caption{
 Diagram contributing to
 the charged lepton $A$-term. $F_N^k$ and $F_L^j$  represent the auxiliary
fields associated with $N_k$ and $L_j$, respectively. The
$\otimes$ at the right-hand side indicates the neutrino $B$-term
insertion while the $\otimes$ at the left-hand side represents the
supersymmetric vertex given by $M_k$.} \label{aell}
\end{figure}

As it is depicted in Fig. \ref{aell}, the neutrino $B$-term can
induce a correction to the $A$-term of charged leptons. If $B_\nu$
is imaginary, the correction which is proportional to $B_\nu$ will
 also be  imaginary, contributing to the EDMs of corresponding
charged lepton: \be \vec d_{\ell_i}=-\frac{2 \alpha}{(4 \pi)^3}
\sum_{\alpha k} \left( \frac{V_{01a}}{c_w} \right) \left(
\frac{V_{01a}}{c_w}+\frac{V_{02a}}{s_w}\right){{\rm Im}[B_\nu]
m_{\ell_i} \over m_a^3} (Y_\nu^{ki})^*Y_\nu^{ki}
f(\frac{m_{\tilde{L}}^2}{m_a^2},\frac{m_{\tilde{E}}^2}{m_a^2})\vec
S \ee where $\vec S$ is the spin of the particle; $V_{0ia}$ and
$m_a$ are respectively the mixing and masses of the neutralinos
and \be f( x_L,x_E)=\frac{1}{2}\frac{1}{x_E-x_L}\left( {1-x_L^2+2
x_L \log x_L \over (1-x_L)^3}-{1-x_E^2 +2 x_E \log x_E \over
(1-x_E)^3} \right).\ee
If we assume that the imaginary $B_\nu$ is the  dominant source of
CP-violation contributing to $d_e$, the present strong bound
\cite{pdg}$^)$ on $d_e$ ($d_e<1.4\times 10^{-27}$)  implies ${\rm
Im}[B_\nu]\sum_i|(Y_\nu)_{ie}|^2/(16
\pi^2)\stackrel{<}{\sim}0.1m_{susy}$. This bound can be improved
significantly in the near future.

Recently, it has been shown that an imaginary $B_\nu$  gives an
imaginary correction to $A_u$, inducing a contribution  to the
EDMs of $d_{Hg}$ and $d_n$, too \cite{padova}$^)$.

In principle, the contribution of the  different CP-violating
phases to EDMs can cancel each other. According to Ref.
 \cite{nocancelation}$^)$, if $\mu$ and $a_0$ are the only sources
of CP-violation, it will not be possible to satisfy the upper
bounds on $d_e$ and $d_{Hg}$ by cancelation scenario and  as a
result the phases of $\mu$ and $a_0$ indeed have to be very small.
Now, if we turn on the imaginary $B_\nu$, there will be enough
parameters to satisfy the experimental bounds even if
$\phi_\mu,\phi_{a_0}\sim 1$. This can have novel experimental
implications in accelerator physics \cite{progress}$^)$.
\section{Conclusions} \label{conclusion}
The condition for the electroweak symmetry breaking implies
$|b_H-B_\nu \mu {\rm  Tr}[Y_\nu^\dagger Y_\nu]/16 \pi^2|\sim
m_{susy}^2/\tan \beta$. Assuming that the other supersymmetry
breaking parameters are all of order of a few hundred GeV, this
puts an upper bound on $B_\nu$ which is stronger than the bound
derived from the radiative correction of the $B$-term to $m_\nu$
\cite{grossman}$^)$. Furthermore, unlike the bound derived in Ref.
\cite{grossman}$^)$, the bound discussed in this paper does not
depend on the values of the right-handed neutrino masses.

Even within this bound, an imaginary $B_\nu$ can induce a
significant contribution to the charged lepton EDMs: \be
d_{\ell_i}\sim 10^{-27} \frac{{\rm Im}[B_\nu]}{ m_{\tilde{L}}}
Y_\nu^2 \left( \frac{200 \ {\rm GeV}}{m_{\tilde{L}}}\right)^2
\frac{m_{\ell_i}}{m_e} \ e\ {\rm cm}. \ee Note that if ${\rm
Im}[B_\nu]\sim m_{susy}$ (as it is expected in the framework of
mSUGRA) this contribution can saturate the present bound on $d_e$
\cite{pdg}$^)$.

\section*{Acknowledgments}
I would like to thank the organizers of the XXXXth Moriond
conference on electroweak interactions and unified theories where
this talk was presented. I also acknowledge the European Union
Programme ``Human Resources and Mobility Activity - Marie Curie
Conferences" for partially supporting my participation in  the
Moriond conference.

\end{document}